\pdfoutput=1 
\documentclass{JINST}

\title{Development of a GEM based detector for the CBM Muon Chamber (MUCH)}

\author{S. Biswas$^{a,b}$\thanks{Corresponding author.}~,
D. J. Schmidt$^a$,
A. Abuhoza$^a$,
U. Frankenfeld$^a$,
C. Garabatos$^a$,
J.~Hehner$^a$,
V. Kleipa$^a$,
T. Morhardt$^a$,
C.J. Schmidt$^a$,
H.R. Schmidt$^c$
and J. Wiechula$^c$\\
\llap{$^a$}GSI Helmholtzzentrum f\"ur Schwerionenforschung GmbH,\\
  Planckstrasse 1, D-64291 Darmstadt, Germany\\
\llap{$^b$}National Institute of Science Education and Research,\\
  Sachivalaya Marg, PO: Sainik School, Bhubaneswar - 751 005, Odisha, India\\
\llap{$^c$}Eberhard-Karls-Universit\"at,\\
  T\"ubingen, Germany\\
E-mail: \email{saikat.ino@gmail.com}}

\abstract{The characteristics of triple GEM detectors have been studied systematically by using cosmic ray muons. The minimum ionizing particle (MIP) spectra has been taken for different GEM voltage setting. Efficiency of GEM detector has been measured for cosmic ray.

At high rate operation of GEMs the value of the protection resistor influences the gain and the stability. This feature has been investigated varying both the rate and the value of the protection resistor. This measurement has been performed using both X-ray generator and Fe$^{55}$ source.

The ageing and long-term stability of GEM based detectors has been studied employing both X-ray generator and Fe$^{55}$ source. The ageing study of one GEM module is performed by using a 8 keV Cu X-ray generator to verify the stability and integrity of the GEM detectors over a longer period of time. The accumulated charge on the detector is calculated from the rate of the X-ray and the average gain of the detector. The details of the measurement and results has been described in this article. 
}


\begin{document}



\section{Introduction}\label{sec:intro}

Gas Electron multipliers (GEM) will be used in the Muon Chamber (MUCH) located downstream of the Silicon Tracking System (STS) of the Compressed Baryonic Matter (CBM) experiment along with other sophisticated detectors at the future Facility for Antiproton and Ion Research (FAIR) in Darmstadt, Germany \cite{FS97,CBM,FAIR}. The CBM experiment at FAIR will use proton and heavy ion beams to study matter at extreme conditions and to explore the QCD phase diagram in the region of high baryon densities \cite{CBM2008}. With CBM we will enter a new era of nuclear matter research by measuring rare diagnostic probes never observed before at FAIR energies, and thus CBM has a unique discovery potential. The focus of the CBM experiment is to detect the rarest probes like Charmonia as a testimony of the early fireball, along with the conventional probes by varying experimental parameters e.g. beam energy or the centrality of the collisions. A very high beam intensity will enable to detect the rare probes which was not possible earlier at similar energies for statistical reasons. In the CBM experiment, particle multiplicities and phase-space distributions, the collision centrality and the reaction plane will be determined. For example, the study of collective flow of charmonium and multi-strange hyperons will shed light on the production and propagation of these rare probes in dense baryonic matter. This will only be possible with the application of advanced instrumentation, including highly segmented and fast gaseous detectors.

In this paper, we would like to present the results of characterization of triple GEM detectors by cosmic ray, long-term test and ageing studies of GEM for CBM.

\subsection{CBM muon chamber detector concept}\label{sec:much}

The measurements of $J/\psi \rightarrow \mu^{+} \mu^{-}$ and low mass vector meson decay $\rho^{0} \rightarrow \mu^{+} \mu^{-}$, $\omega^{0} \rightarrow \mu^{+} \mu^{-}$ and $\phi^{0} \rightarrow \mu^{+} \mu^{-}$ in Au + Au collisions at 25 AGeV have been proposed as a key probe to the indication of in-medium modification of hadrons, chiral symmetry restoration and deconfinement at high baryon density $\rho_{b}$.

This measurement would require a sophisticated muon detector (MUCH) located downstream of the Silicon Tracking System (STS) of the CBM experiment. The current design of the muon system consists of 6 iron absorber layers of thickness 20, 20, 20, 30, 35, 100 cm respectively and 18 detector layers with $\sim$ 100 $\mu$m position resolution divided into 6 stations. The muon detector in the CBM experiment will be constructed in such a way that there will be micro-pattern gaseous detectors with high rate capability at least in the first 4 stations and other detectors like pad chambers, thick GEM or straw tubes in the later stations \cite{APFAIR}. A schematical view of the CBM detector with its MUCH detection system is shown in Figure~\ref{CBM}.

\begin{center}
\begin{figure}[tbp]
\centerline{\includegraphics[scale=0.4]{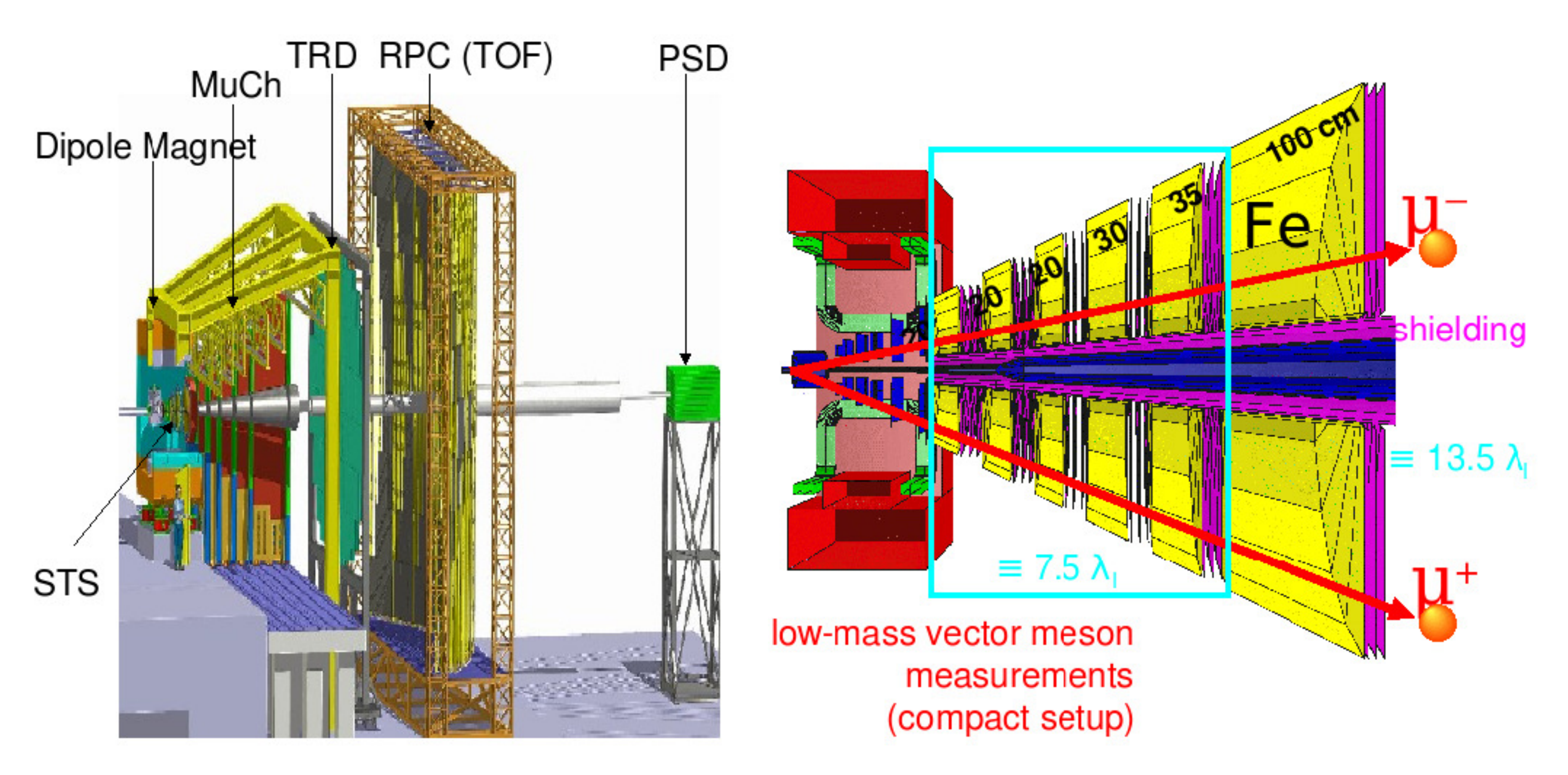}}
\caption{\label{CBM} Schematic view of the CBM experiment: Muon set up (left). Implementation of the muon detection system in GEANT (right).}\label{CBM}
\end{figure}
\end{center}

Simulation studies of the CBM detector system yield that in the first GEM layer of the muon chambers hit densities (H) of 0.5 hits/cm$^{2}$/event at an event rate (R) of 10$^{7}$ Hz will occur. The detectors will have to withstand a large accumulated charge. If the gas gain (G) of the detector is 10$^{3}$ and the number of primary electrons/track (P) is 30 (for Argon and CO$_{2}$) then after amplification the total number of electrons (N$_{e}$) is 1.5 $\times$ 10$^{11}$ /cm$^{2}$/s [N$_{e}$=H$\times$R$\times$P$\times$G]. The accumulated charge per year (Q$_{y}$) computes to around 0.75 C/cm$^{2}$. In the 10 years period of the CBM experiment the total accumulated charge will thus be 7.5 C/cm$^{2}$.  Ageing studies are therefore crucial for extended efficient detector operation.

\section{Cosmic ray test set-up}\label{sec:cosmicsetup}

Characteristics of GEM detectors have been studied systematically by using cosmic ray muons. The minimum ionizing particle (MIP) spectra has been taken for different GEM voltage setting. The drift, transfer and induction gaps of the detector are 2 mm, 2 mm and 2 mm respectively. The voltage to the drift plane and individual GEM plates has been applied by 7 channel power supply (HVG210). Although there is a segmented readout pad (256 pads of 6$\times$6~mm$^2$ area) the signal in this study was obtained from two add up boards (named as S0 and S1 and each board summed 128 pads) and a single input from each board is fed to a charge sensitive preamplifier. After that a PXI LabVIEW based data acquisition system is used. The drift field, induction field and the two-transfer fields were kept constant at 2.5~kV/cm, 2~kV/cm and 3~kV/cm respectively. The detector has been operated with argon and CO$_{2}$ in 70/30 mixing ratio. The pulse height distribution for the GEM has been studied by a NI PXI-5105 sampling ADC. In the cosmic ray measurement the trigger of the sampling ADC has been taken by the coincidence signal of three plastic scintillators. The cosmic ray test set-up along with the GEM detector is shown in Figure~\ref{cosmic_setup}.

\begin{center}
\begin{figure}[tbp]
\centerline{\includegraphics[scale=0.4]{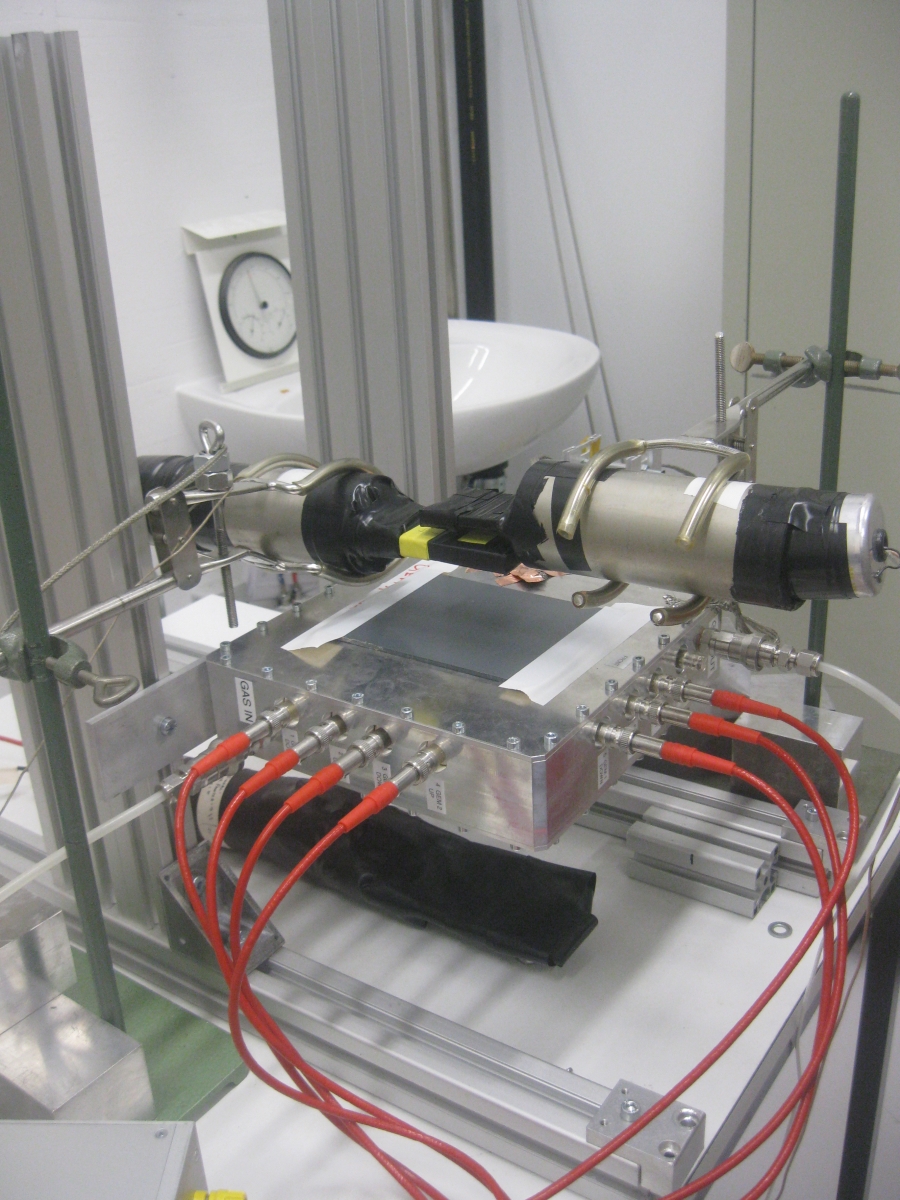}}
\caption{\label{cosmic_setup} Cosmic ray test set-up.}\label{cosmic_setup}
\end{figure}
\end{center}

\subsection{Results of cosmic ray test}\label{sec:cosmicresults}
Pulse height distribution is studied and the efficiency and crosstalk is measured by using cosmic ray. A typical minimum ionizing particle (MIP) spectrum at GEM voltage 400-395-300 V is shown in Figure~\ref{mip}. In this spectrum the signal and noise peak are well separated. The spectrum is fitted with a Landau distribution. The overflow at ADC channel 700 is due to saturation of the pre-amplifier.

\begin{center}
\begin{figure}[tbp]
\centerline{\includegraphics[scale=0.6]{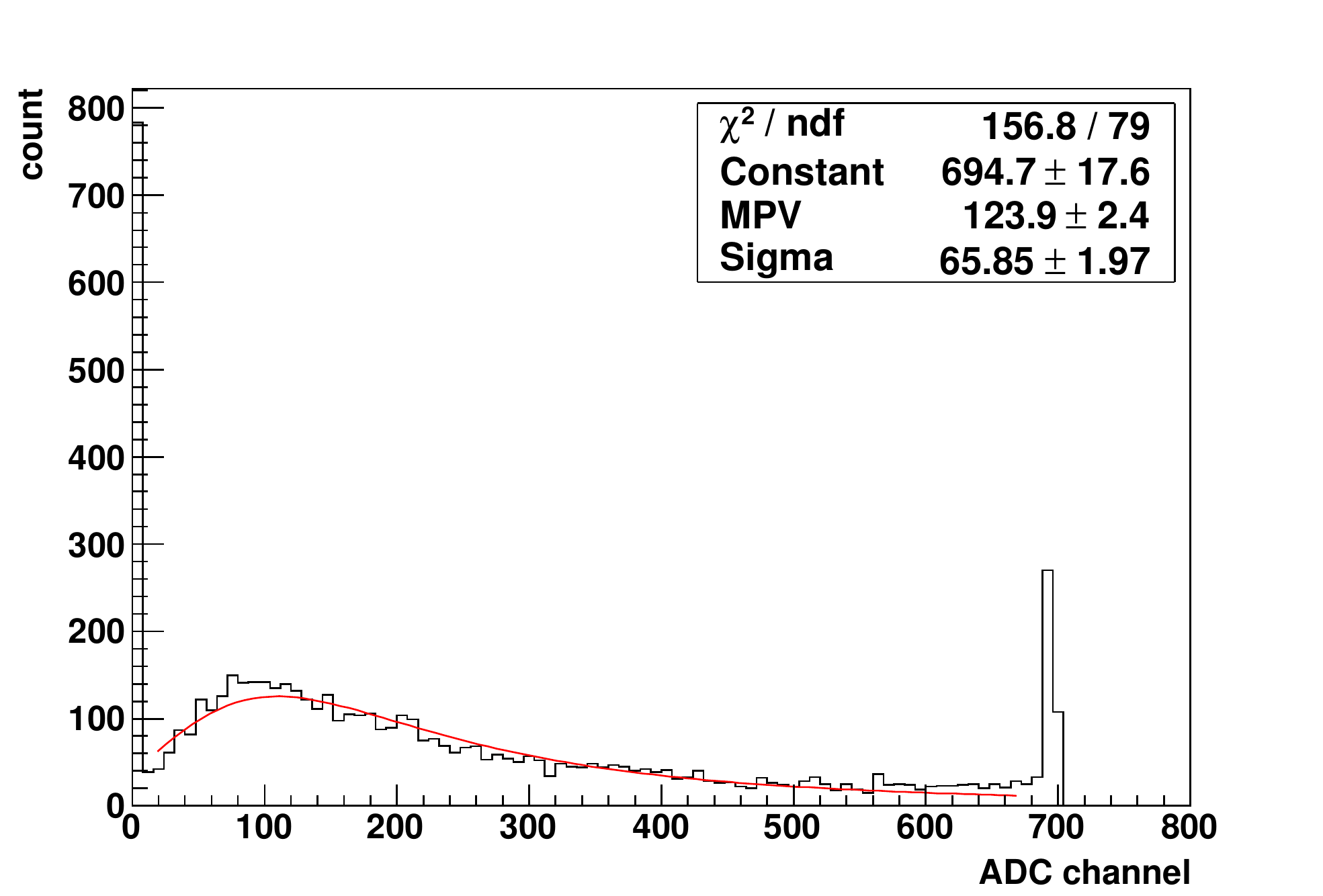}}
\caption{\label{mip} A typical MIP spectrum at GEM voltage 400-395-390 V.}\label{mip}
\end{figure}
\end{center}

The MPV of the Landau fitting for minimum ionizing particles are measured at different GEM voltage settings. The MPV as a function of global GEM voltage is plotted in Figure~\ref{mpv}.

\begin{center}
\begin{figure}[tbp]
\centerline{\includegraphics[scale=0.6]{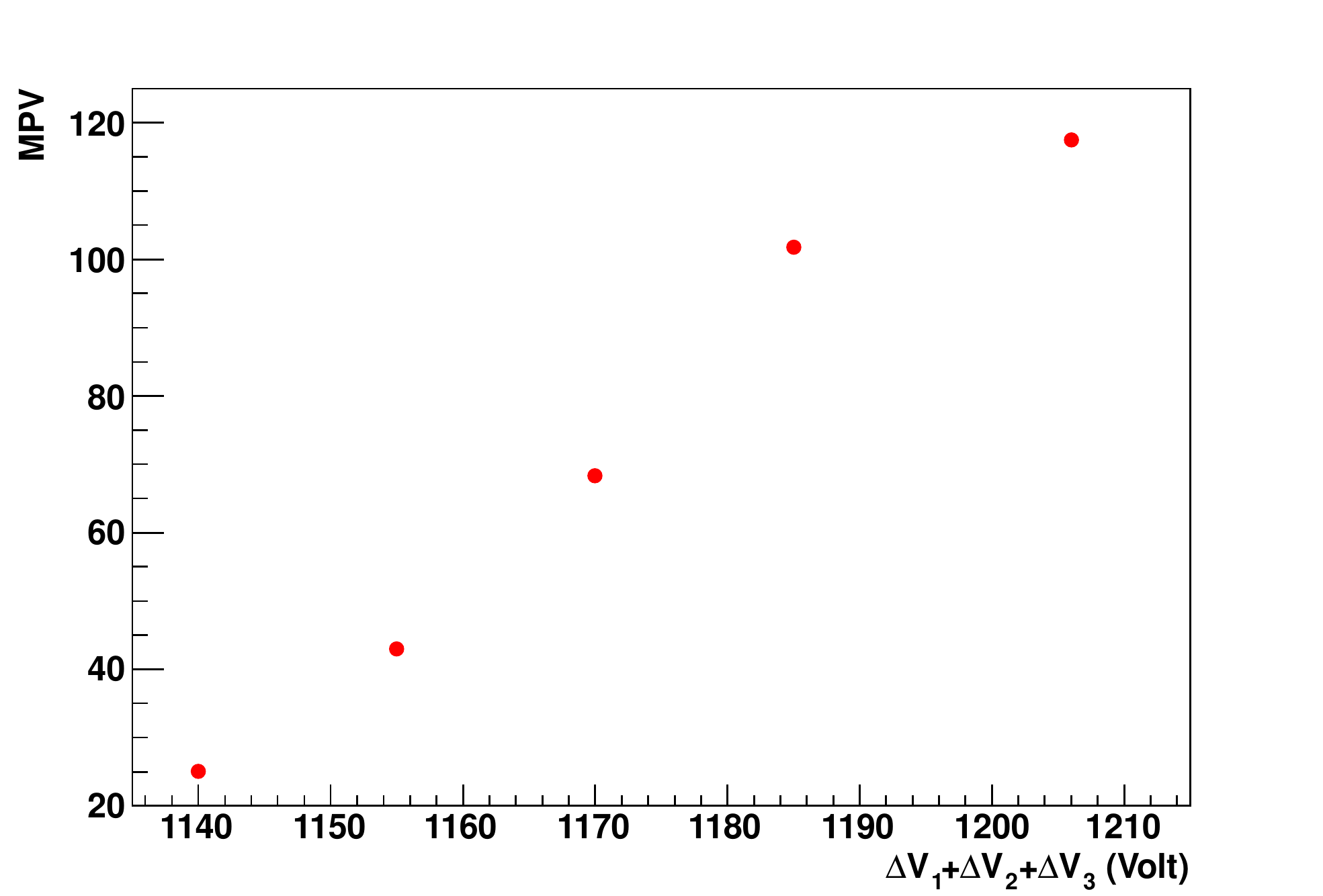}}
\caption{\label{mpv} MPV of Landau fitting as a function of global GEM voltage.}\label{mpv}
\end{figure}
\end{center}

The efficiency of the GEM detector is measured in the same cosmic ray test bench by varying the global GEM voltage. The efficiency as a function of the global GEM voltage is shown in Figure~\ref{eff}. A plateau at 95\% has been achieved for cosmic ray.

\begin{center}
\begin{figure}[tbp]
\centerline{\includegraphics[scale=0.6]{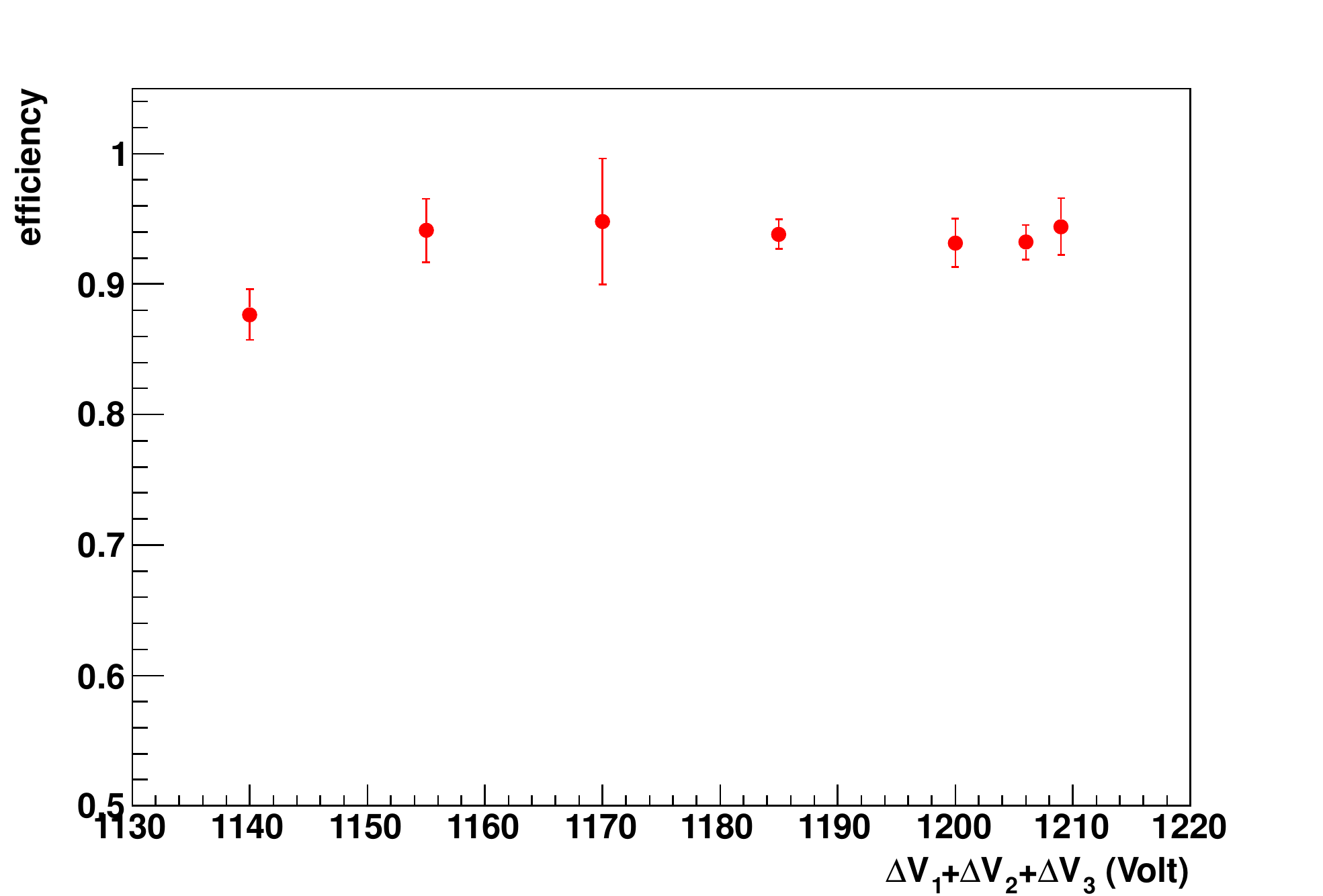}}
\caption{\label{eff} Efficiency as a function of the global GEM voltage.}\label{eff}
\end{figure}
\end{center}

The total number of read out pad for the GEM was 256 and of area 6$\times$6~mm$^2$ as mentioned in section~\ref{sec:cosmicsetup}. The readout pads are connected to two sum-up boards each with 128 pins. With the scintillator trigger both the signals have been read out. About 10\% crosstalk in the pad plane has been observed between the two segments of the GEM.

\section{Effect of protection resistor}\label{sec:prores}

At high rate operation of GEMs the value of the protection resistor influences the gain. This feature has been investigated varying both the rate of X-ray and the value of the protection resistor. This measurement has been performed using both Fe$^{55}$ source and Cu X-ray generator. 

In this whole set of measurements the GEM voltages i.e. $\Delta$V$_{1}$, $\Delta$V$_{2}$, $\Delta$V$_{3}$ have been kept at 400~V, 395~V and 390~V and the drift, induction and two-transfer fields were kept constant at 2.5~kV/cm, 2~kV/cm and 3~kV/cm respectively. The gain of the detector has been expressed by obtaining the mean position of 5.9 keV peak of Fe$^{55}$ X-ray spectrum with Gaussian fitting \cite{CBM11,SB11}. Using  Fe$^{55}$ source the rate has been varied in two different ways. The collimator has been fixed with the detector and the distance of the source from the detector is varied and the diameter of the collimator is varied keeping the source to detector distance constant. For these two sets of measurements the variations of the mean ADC channel of 5.9 keV peak of Fe$^{55}$ X-ray spectrum, as a function of the rate are shown in Figure~\ref{coldet} and Figure~\ref{coldia} respectively. For each set of measurements three sets of protection resistance value have been chosen. The sets are 11~M$\Omega$ protection resistance to the drift plane and both the planes (top and bottom) of all three GEM foils, 1~M$\Omega$ protection resistance to the drift plane and both the planes of all three GEM foils and 11~M$\Omega$ protection resistance to the drift plane and on the top plane of all three GEM foils.
\begin{center}
\begin{figure}[tbp]
\centerline{\includegraphics[scale=0.6]{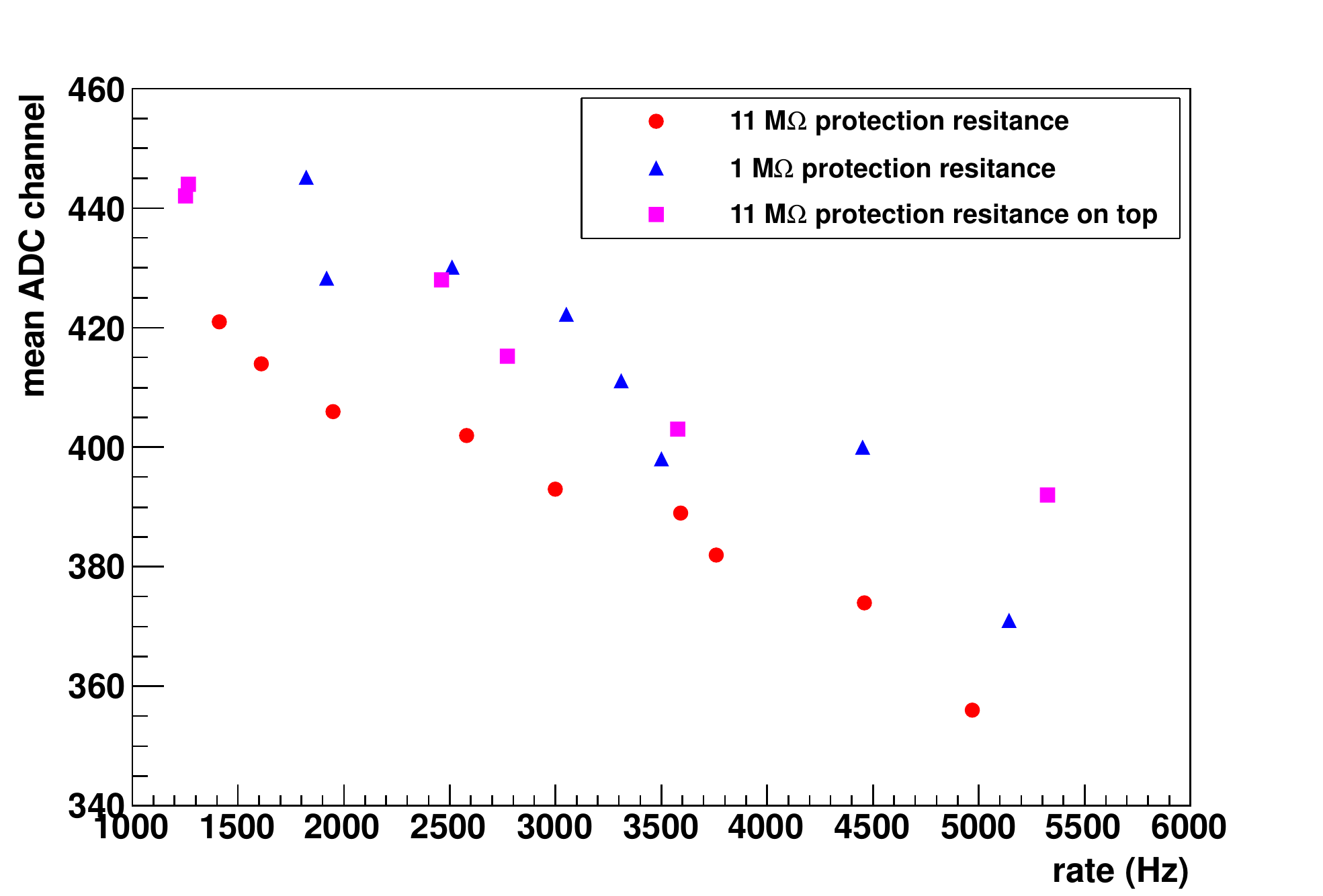}}
\caption{\label{coldet} Variation of the mean ADC channel (mean position of 5.9 keV peak of Fe$^{55}$ X-ray spectrum with Gaussian fitting) as a function of the rate varying the detector and source distance with collimator fixed with the detector.}\label{coldet}
\end{figure}
\end{center}
\begin{center}
\begin{figure}[tbp]
\centerline{\includegraphics[scale=0.6]{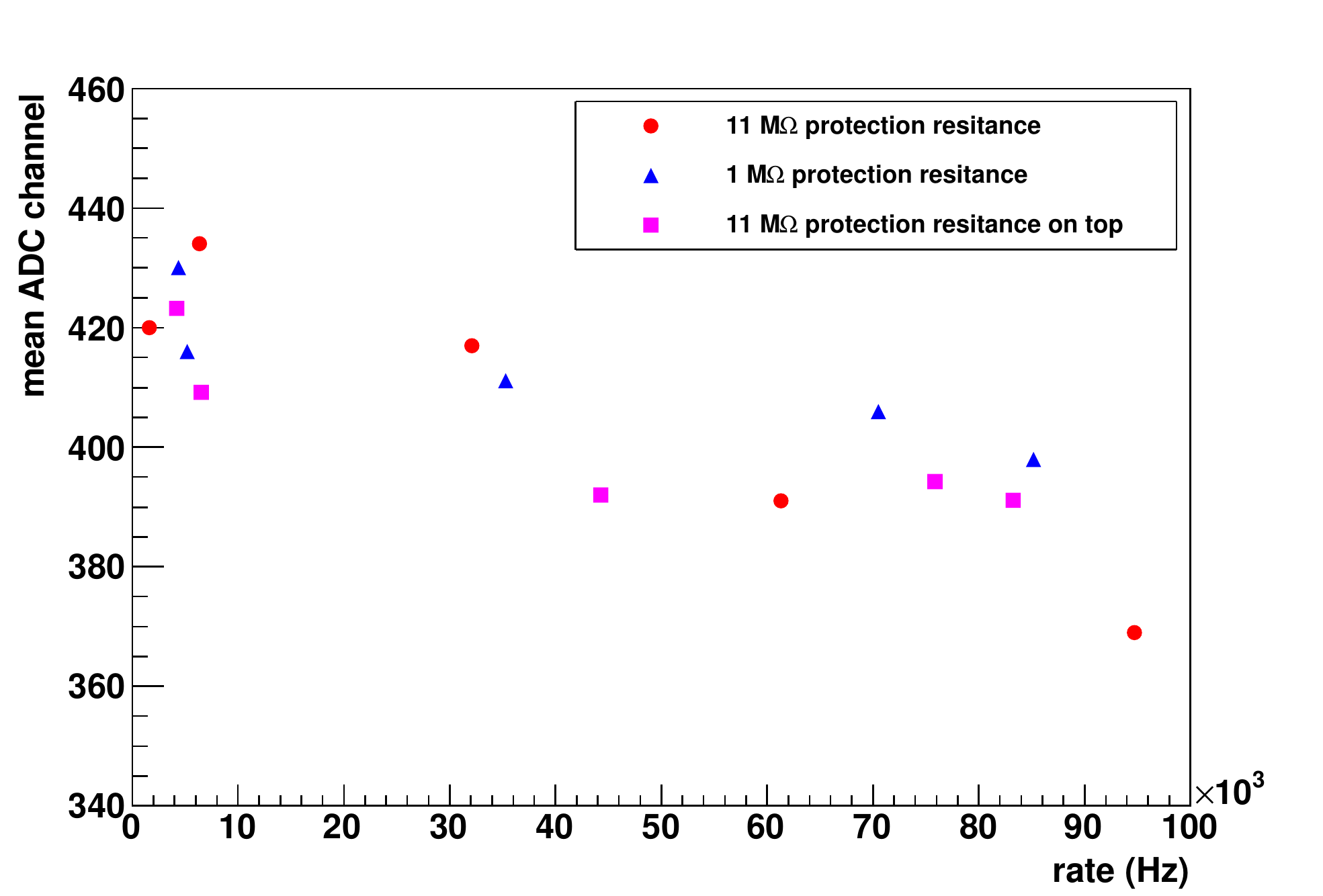}}
\caption{\label{coldia} Variation of the mean ADC channel (mean position of 5.9 keV peak of Fe$^{55}$ X-ray spectrum with Gaussian fitting) as a function of the rate varying the diameter of the collimator.}\label{coldia}
\end{figure}
\end{center}

In both the set of measurements the mean ADC channel of 5.9 keV peak of Fe$^{55}$ X-ray spectrum decreases with the rate in all three sets of protection resistance value. The rate of decrease of the mean is maximum with 11~M$\Omega$ protection resistance to both the planes of all three GEM foils.

\begin{center}
\begin{figure}[tbp]
\centerline{\includegraphics[scale=0.32]{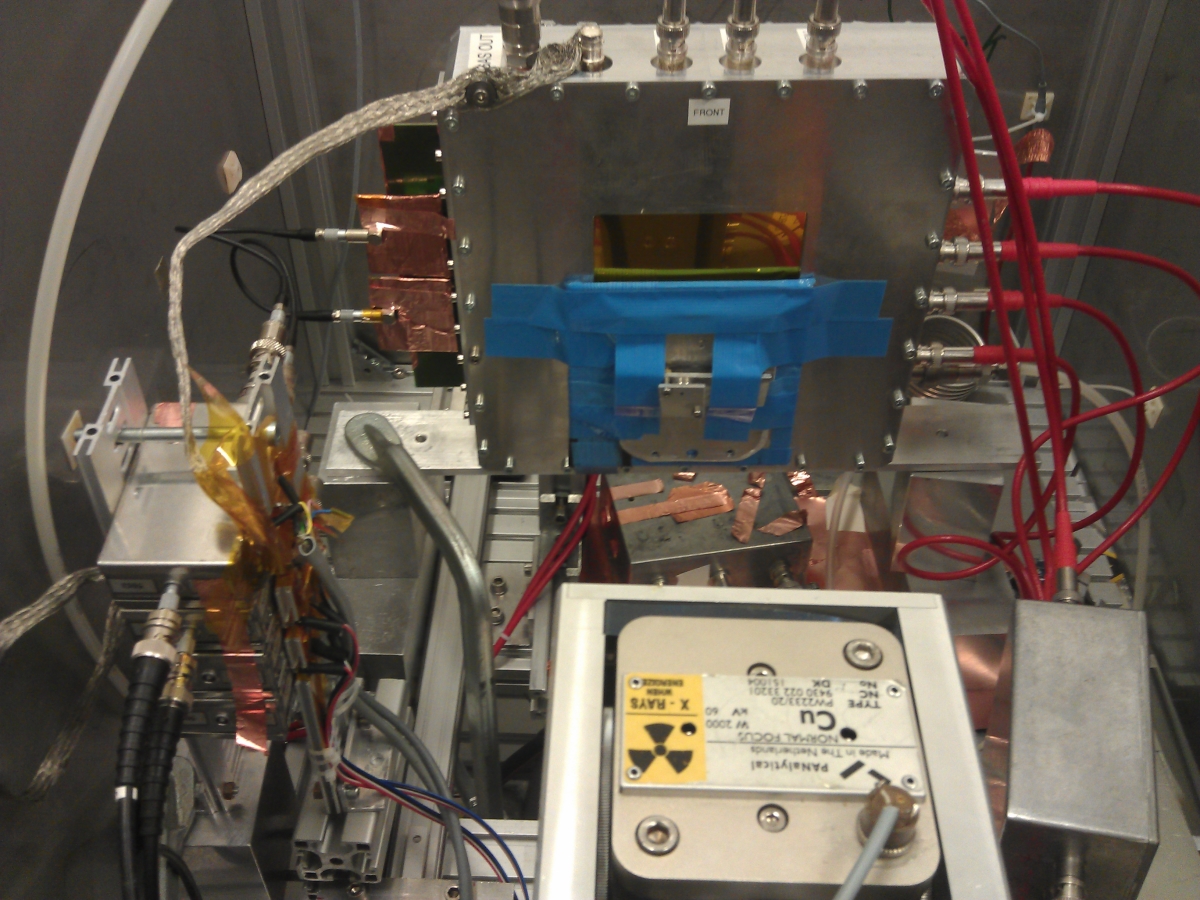}}
\caption{\label{gen_setup} Set-up with Cu X-ray generator.}\label{gen_setup}
\end{figure}
\end{center}

\begin{center}
\begin{figure}[tbp]
\centerline{\includegraphics[scale=0.6]{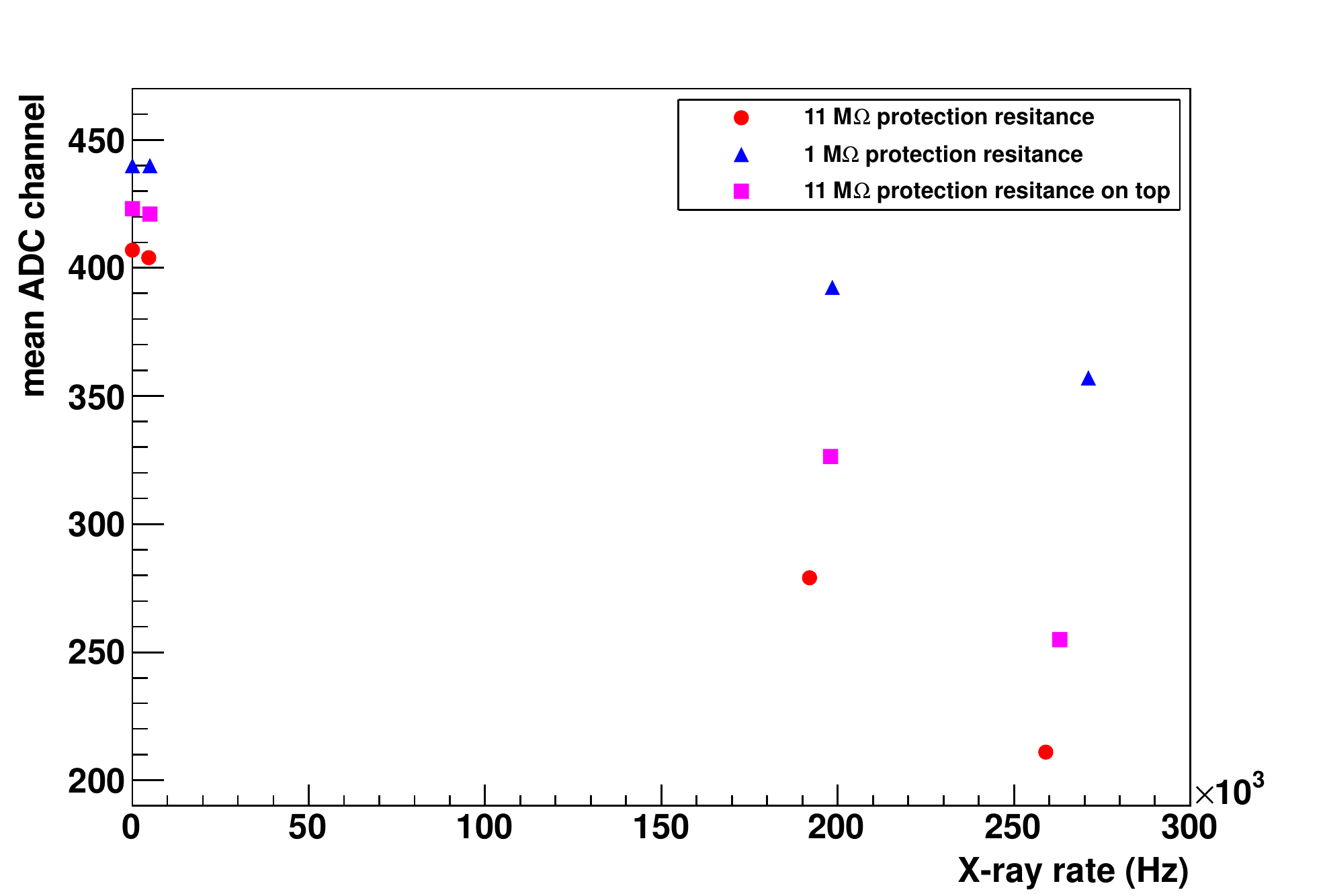}}
\caption{\label{gen} Variation of the mean ADC channel (mean position of 5.9 keV peak of Fe$^{55}$ X-ray spectrum with Gaussian fitting) as a function of the rate using Cu X-ray generator.}\label{gen}
\end{figure}
\end{center}

Dependence of gain on protection resistance value is also measured using the Cu X-ray generator. The upper part of the GEM is exposed to the X-ray from the generator and the Fe$^{55}$ source is placed at the bottom part of the detector. The Fe$^{55}$ spectrum is taken from the bottom part of the GEM at the same time when the upper part of the detector is exposed to the Cu X-ray from the generator. The Cu X-ray generator along with the GEM detector is shown in Figure~\ref{gen_setup}. In this case the rate of X-ray from the generator is changed by varying the bias voltage. 

The variations of the mean ADC channel of 5.9 keV peak of Fe$^{55}$ X-ray spectrum, as a function of the Cu X-ray rate is shown in Figure~\ref{gen} for all three sets of protection resistance combination i.e. 11~M$\Omega$ to the drift plane and both the planes (top and bottom) of all three GEM foils, 1~M$\Omega$ to the drift plane and both the planes of all three GEM foils and 11~M$\Omega$ to the drift plane and on the top plane of all three GEM foils. It is clear from the Figure~\ref{gen} that in all three combination of protection resistance value the mean decreases with the rate.

\section{Long-term study of GEM}\label{sec:long}

The long-term stability of the GEM has been studied using the same test setup. One GEM detector has been tested for about 14 days continuously. The voltages to the GEM 1, 2 and 3 were 400~V, 395~V and 390~V whereas the drift, induction and the two-transfer fields were kept constant at 2.5~kV/cm, 2~kV/cm and 3~kV/cm respectively. A collimated Fe$^{55}$ source is placed in front of the detector and the X-ray spectra have been taken for 10-minute interval. The mean position of 5.9 keV Fe$^{55}$ X-ray peak has been recorded with time. The variation of mean as a function of time is shown in Figure~\ref{meantime}. The ambient temperature and pressure also recorded continuously and the T/p as a function of time is shown in Figure~\ref{tbyptime}. A definite correlation between mean and T/p has been observed which is well known for any gas detector.
\begin{center}
\begin{figure}[tbp]
\centerline{\includegraphics[scale=0.6]{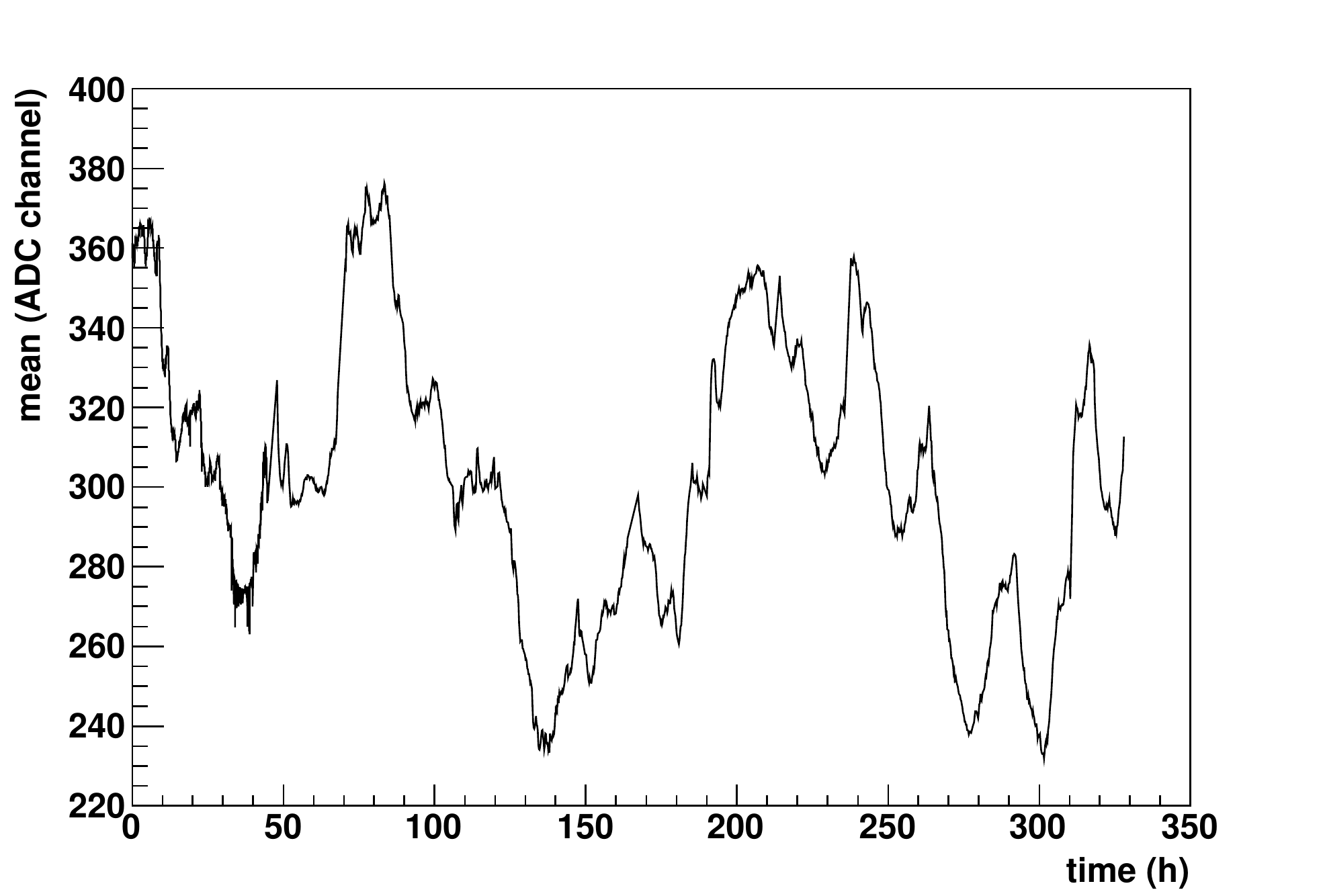}}
\caption{\label{meantime} Mean (mean position of 5.9 keV peak of Fe$^{55}$ X-ray spectrum with Gaussian fitting) as a function of the period of operation.}\label{meantime}
\end{figure}
\end{center}
\begin{center}
\begin{figure}[tbp]
\centerline{\includegraphics[scale=0.6]{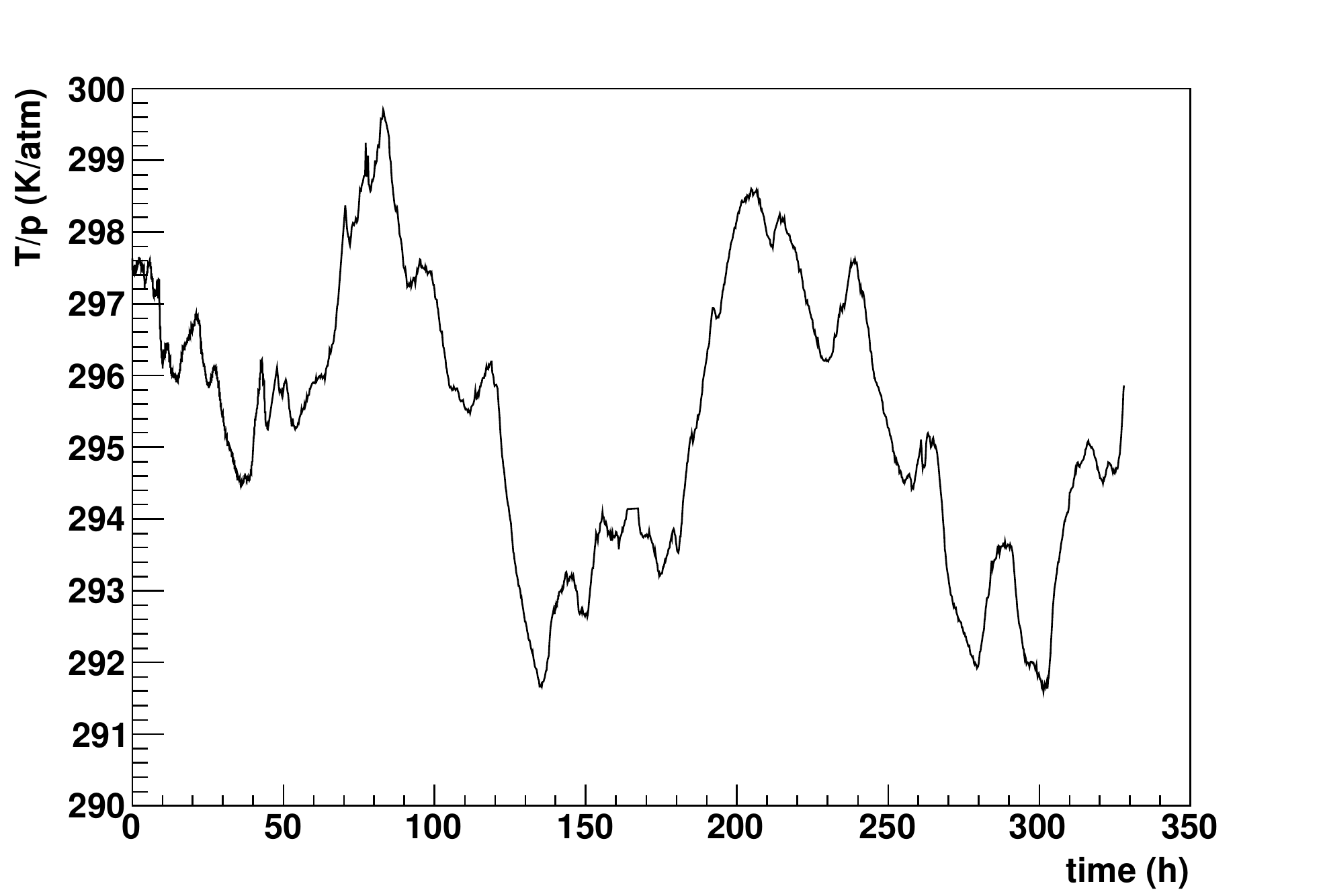}}
\caption{\label{tbyptime} T/p as a function of the period of operation.}\label{tbyptime}
\end{figure}
\end{center}

Therefore, a corrected and normalized gain g was computed from the effective gain G according to

\begin{equation}\label{eq1}
g={G\over Ae^{BT/p}}
\end{equation}

The T/p is the ratio of temperature and pressure. A and B are fit parameters, determined by fitting the exponential function,

\begin{equation}\label{eq2}
G(T/p)={Ae^{BT/p}}
\end{equation}

to the correlation plot of Figure~\ref{fit}.
\begin{center}
\begin{figure}[tbp]
\centerline{\includegraphics[scale=0.6]{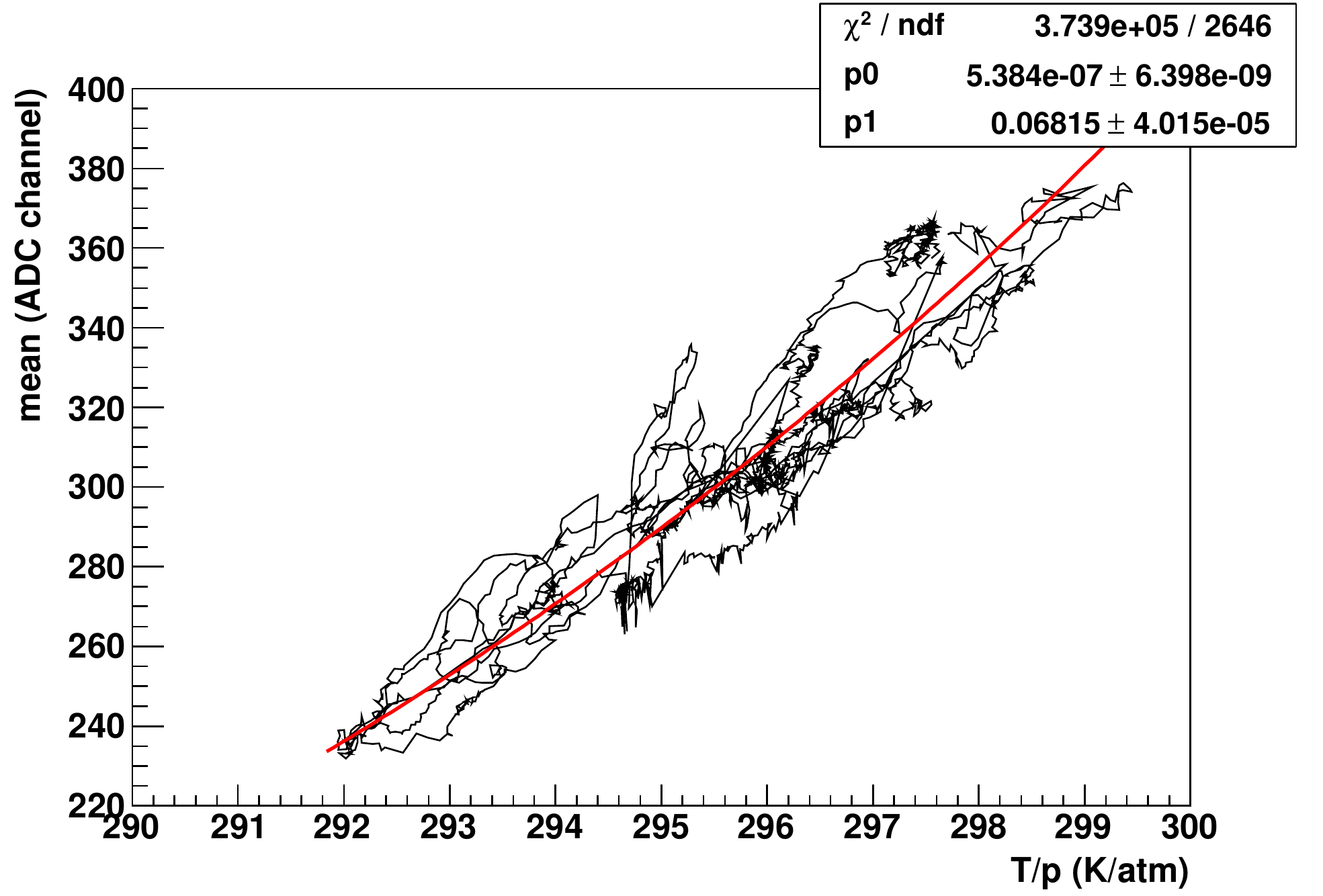}}
\caption{\label{fit} Correlation of detector gain and temperature \& pressure: pulse height versus T/p and the exponential fits.}\label{fit}
\end{figure}
\end{center}

The exponential dependence in Eq.~\ref{eq2} is deduced by assuming inverse proportionality of the Townsend coefficient $\alpha$ to the mass density $\rho$; and thus $\alpha$ $\propto$ 1/$\rho$ $\propto$ T/p \cite{MCA03}.
\begin{center}
\begin{figure}[tbp]
\centerline{\includegraphics[scale=0.6]{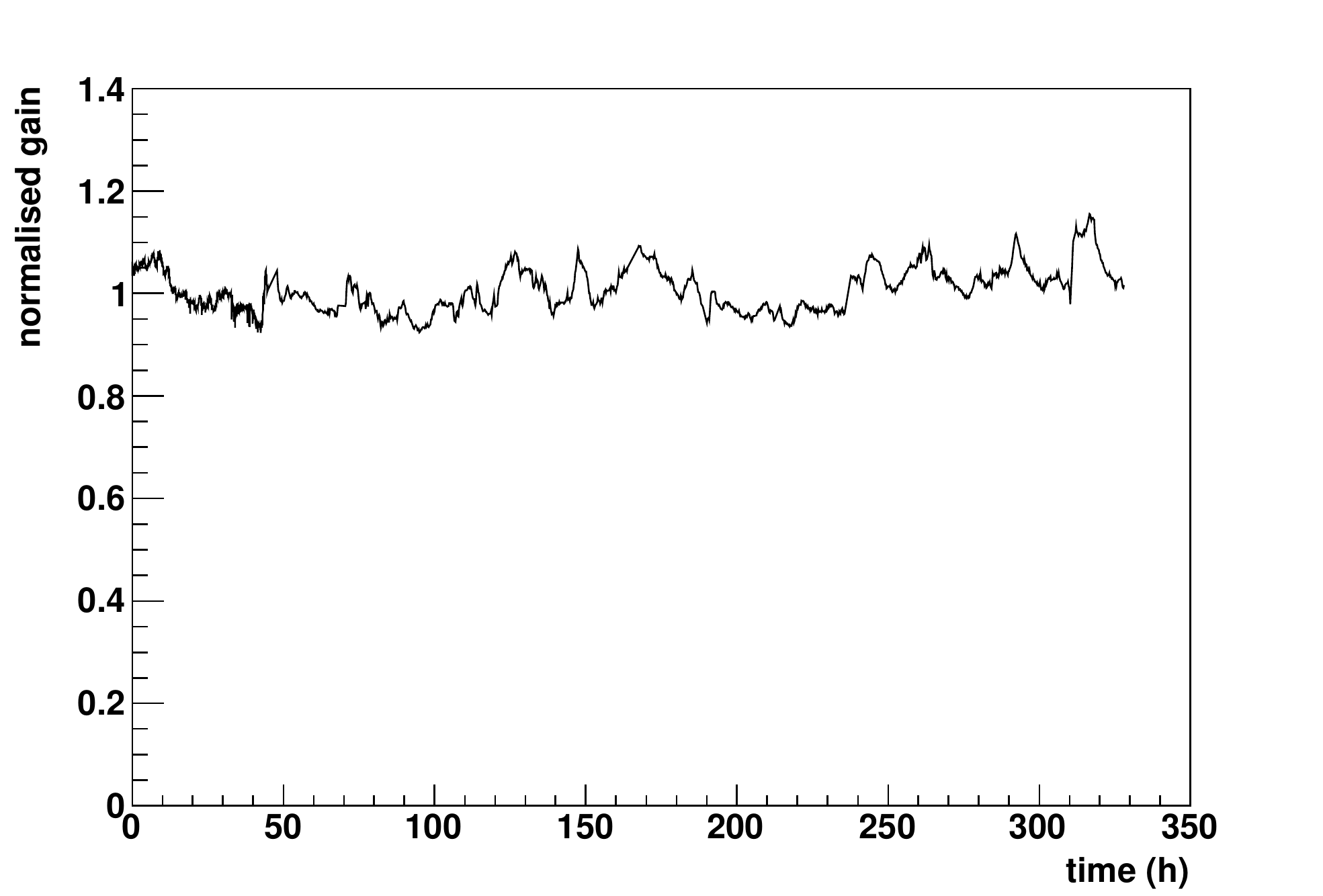}}
\caption{\label{gaintime} Normalized gain as a function of the period of operation.}\label{gaintime}
\end{figure}
\end{center}

The variation of the normalized gain as obtained by Eq.~\ref{eq1} is plotted in Figure~\ref{gaintime}. Still there is about 10\% variation of the gain from the average value with time. This variation might arises from the variation of the  O$_2$ concentration in the gas as well as variation of the gas ratio due to changes of the characteristics of the mass flow controller with temperature.

The variation of energy resolution with time for the detector is shown in Figure~\ref{resotime}. The energy resolution varies between 17-20\% in the whole period of operation. 

\begin{center}
\begin{figure}[tbp]
\centerline{\includegraphics[scale=0.6]{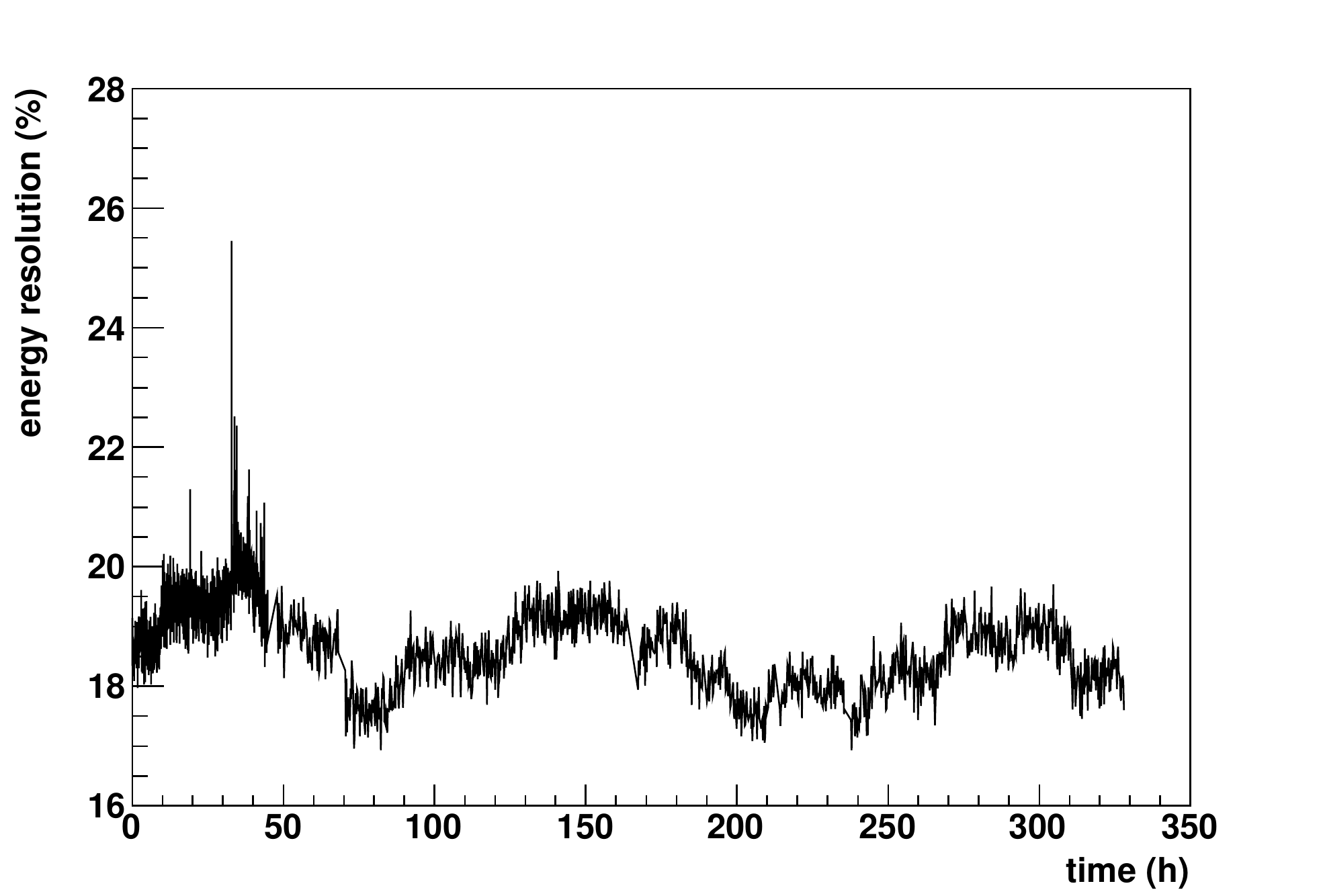}}
\caption{\label{resotime} Energy resolution as a function of the period of operation.}\label{resotime}
\end{figure}
\end{center}

\section{Preliminary ageing study of GEM}\label{sec:ageing}

An infrastructure has been set up at the GSI detector laboratory to study the ageing properties of gas filled detectors such as multi wire proportional chambers (MWPC), gas electron multipliers (GEM) etc \cite{AA12,AA13}. An accuracy in the relative gain measurement below 1\% has been achieved by monitoring environmental conditions and by systematic improvements of the measuring equipment \cite{AA13}. The ageing study of one GEM module is performed in the same ageing measurement set-up by using a 8 keV Cu X-ray generator to verify the stability and integrity of the GEM detectors over a period of time. The GEM has been operated at 395-390-385 V with drift, induction and transfer field at 2.5~kV/cm, 2~kV/cm and 3~kV/cm respectively. The centre of the upper part of the GEM (region A) was exposed to high rate Cu X-rays for 10 minutes. Subsequently the Fe$^{55}$ spectra have been observed for 1 minute each from upper and lower part (region~B) of the GEM. The Fe$^{55}$ source has been placed in such a way that the Fe$^{55}$ X-rays direct toward the upper part of the GEM in the same spot, which was exposed by Cu X-ray. The ratio of the mean position of 5.9 keV Fe$^{55}$ X-ray peak from upper side and lower side of GEM is the normalized gain and corrects the effect of pressure and temperature variations. The whole measurement is performed for about 70 hours. The schematical view of the ageing measurement setup is shown in Figure~\ref{ageingschema}. The rate of the Cu X-ray was about 240~kHz. The rate of X-ray on two sides of the detector as a function of time is shown in Figure~\ref{ratetime}.

\begin{center}
\begin{figure}[tbp]
\centerline{\includegraphics[scale=0.6]{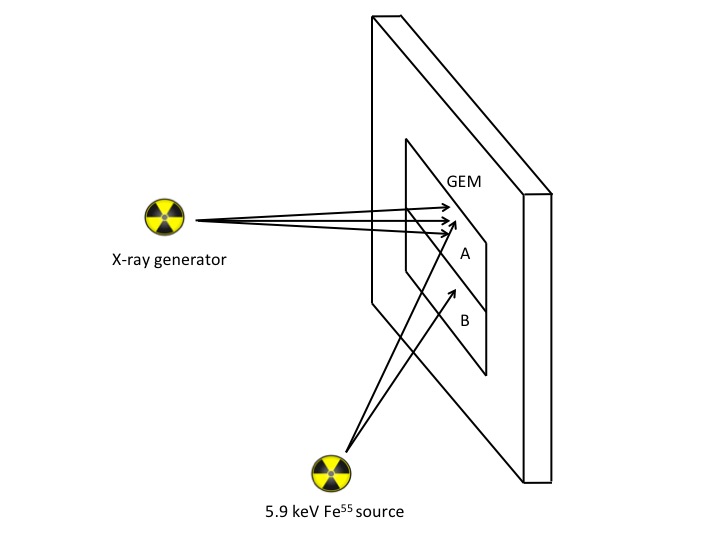}}
\caption{\label{ageingschema} Schematical view of the ageing measurement set-up.}\label{ageingschema}
\end{figure}
\end{center}
\begin{center}
\begin{figure}[tbp]
\centerline{\includegraphics[scale=0.6]{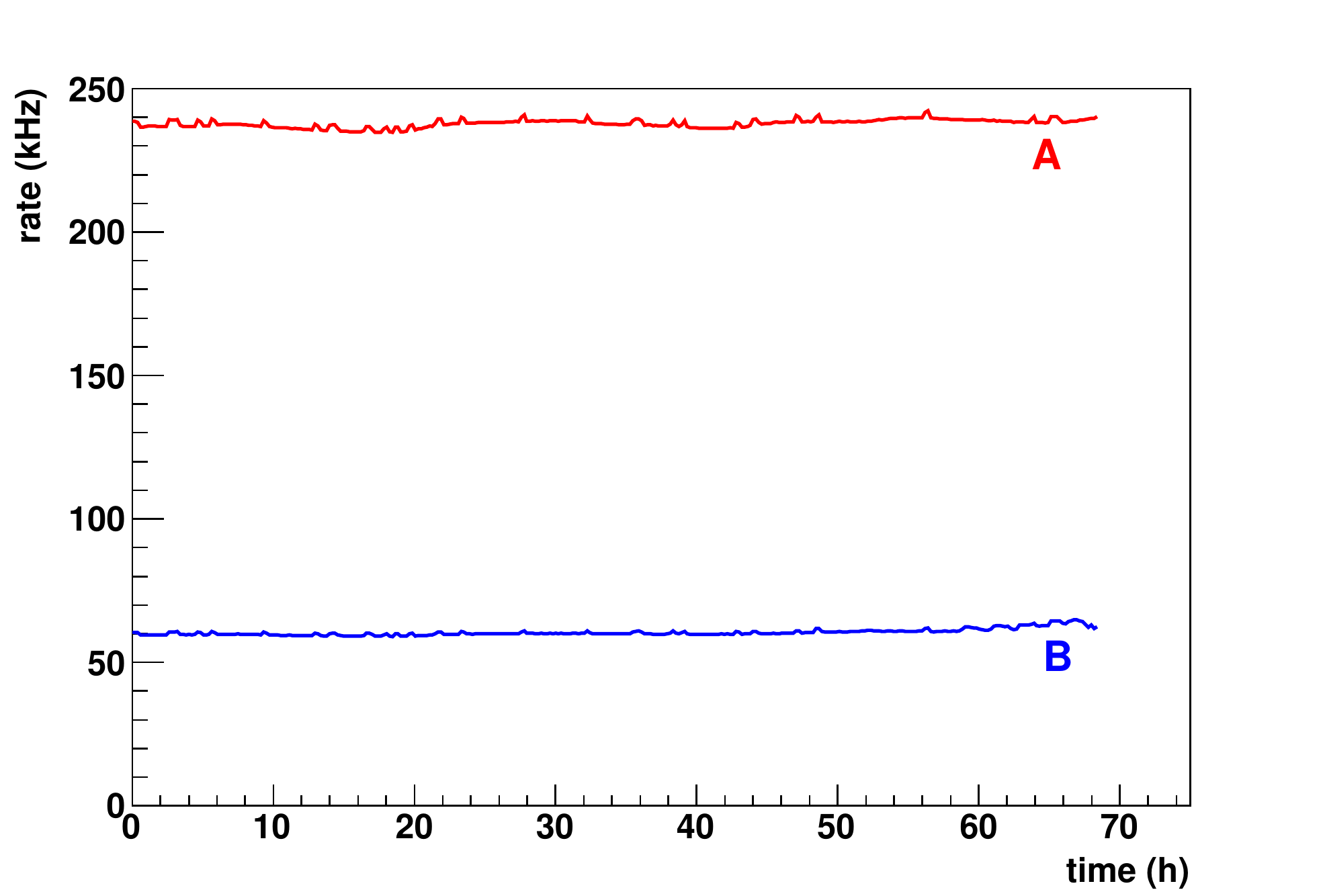}}
\caption{\label{ratetime} Rate as a function of time.}\label{ratetime}
\end{figure}
\end{center}
\begin{center}
\begin{figure}[tbp]
\centerline{\includegraphics[scale=0.6]{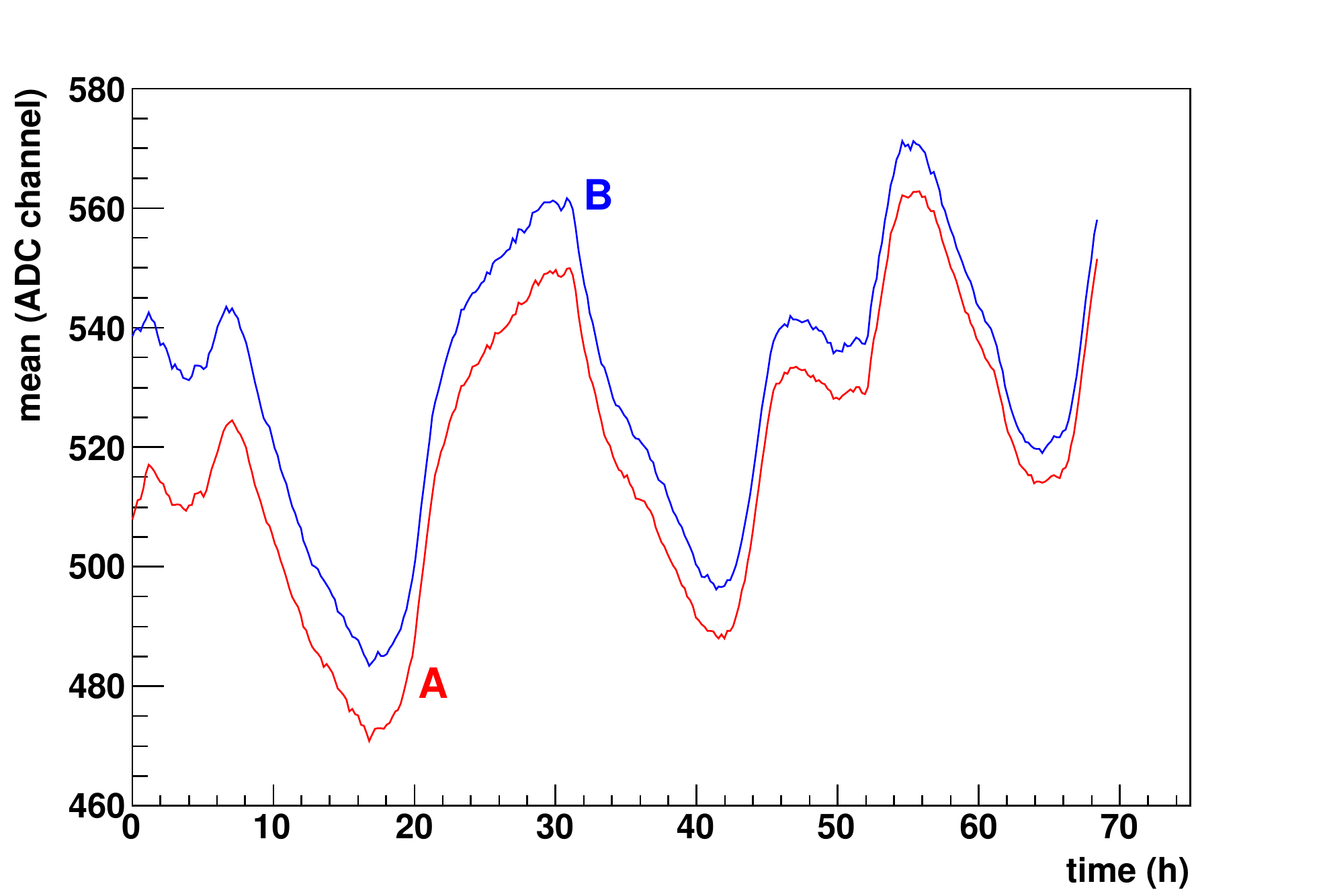}}
\caption{\label{mean_time} Mean (mean position of 5.9 keV peak of Fe$^{55}$ X-ray spectrum with Gaussian fitting) as a function of time.}\label{mean_time}
\end{figure}
\end{center}

The mean Fe$^{55}$ peak positions obtained from the region A and B as a function of time is shown in Figure~\ref{mean_time}. The mean varies with time due to change of pressure and temperature. The gain from the two regions are slightly different due to the difference in the gain of the pre-amplifiers.

\begin{center}
\begin{figure}[tbp]
\centerline{\includegraphics[scale=0.6]{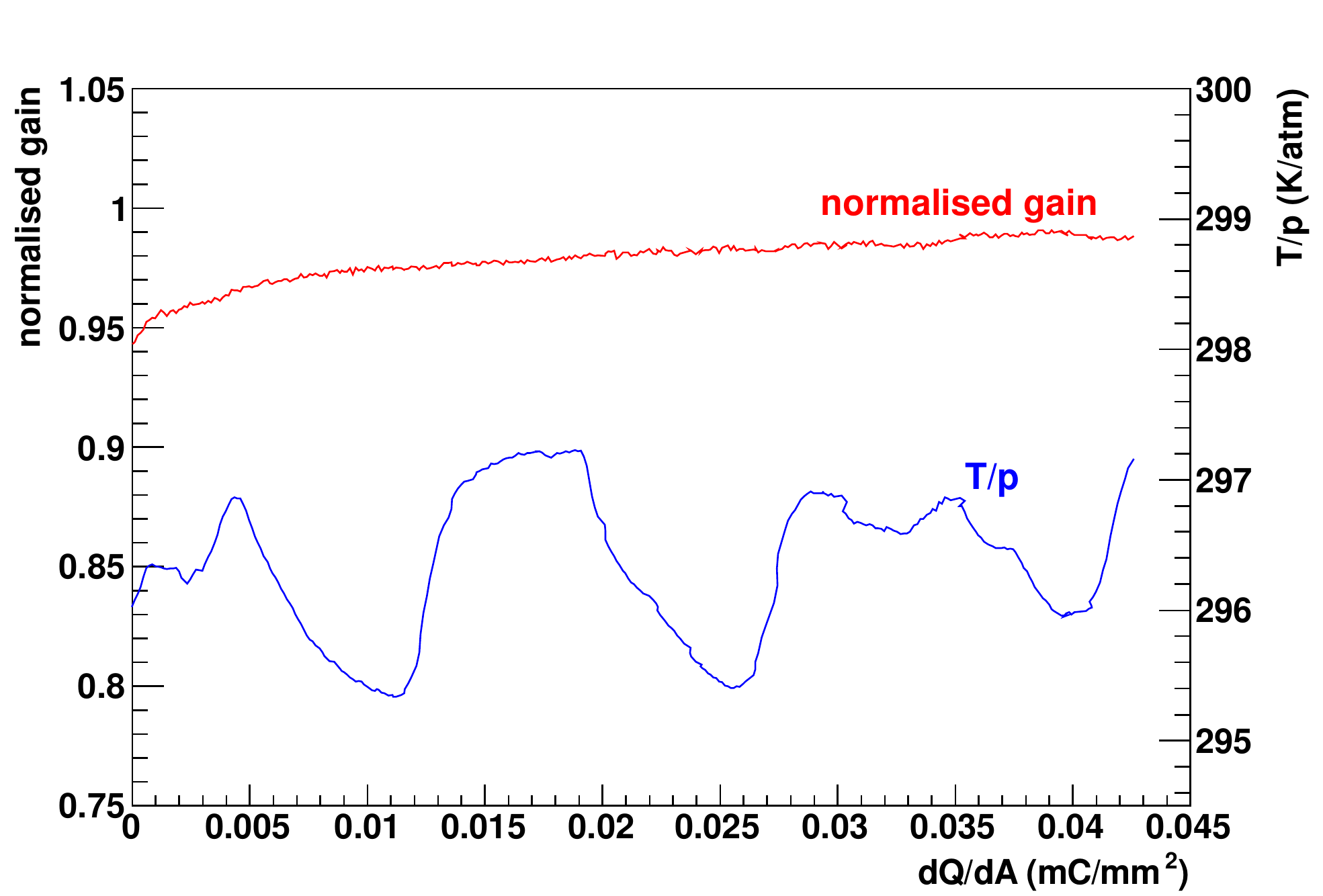}}
\caption{\label{gain_charge} The normalized gain and the T/p as a function of accumulated charge per unit area.}\label{gain_charge}
\end{figure}
\end{center}
\begin{center}
\begin{figure}[tbp]
\centerline{\includegraphics[scale=0.6]{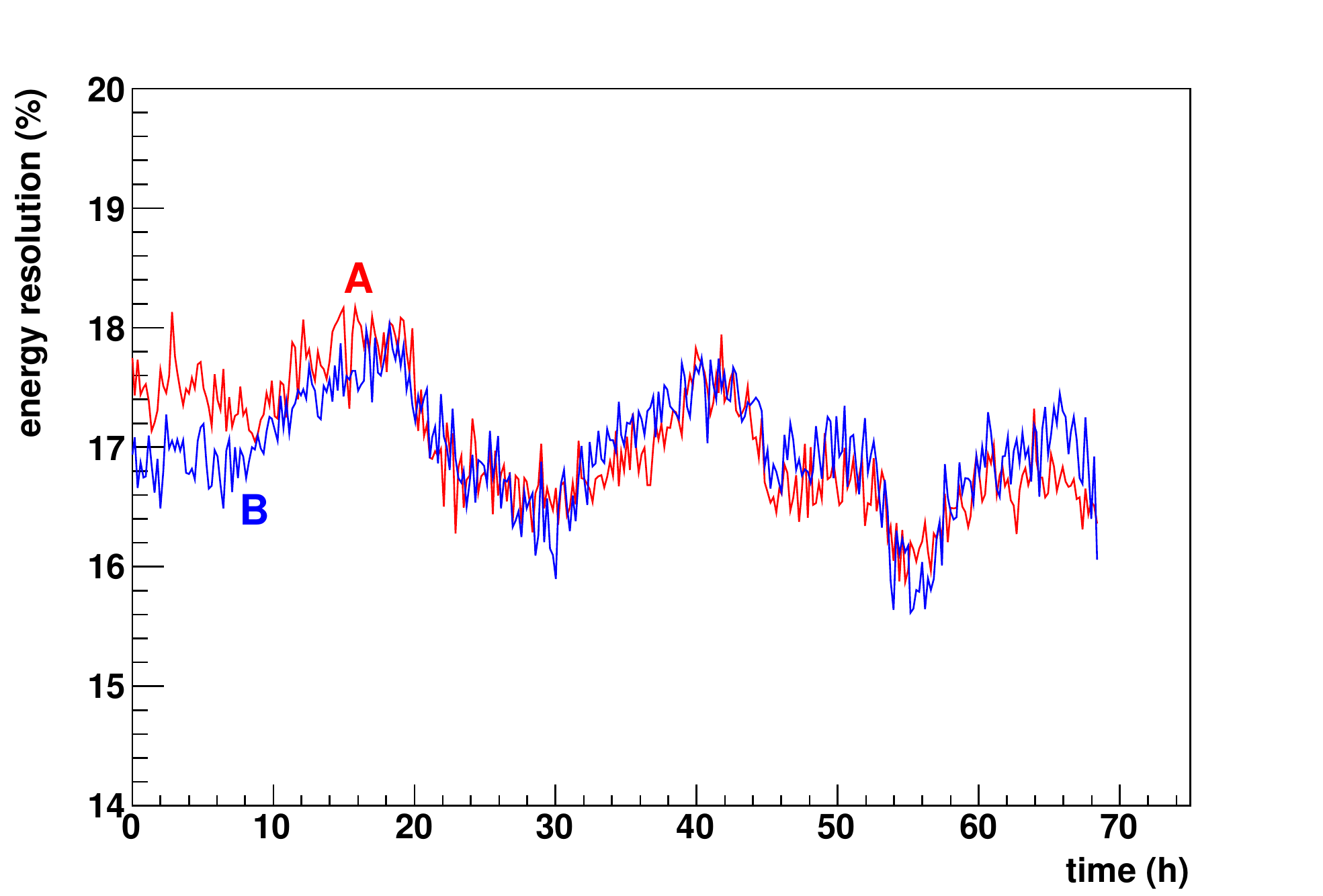}}
\caption{\label{resotimeageing} Energy resolution as a function of time.}\label{resotimeageing}
\end{figure}
\end{center}

The accumulated charge on the detector is calculated from the rate of the X-ray and the average gain of the detector. The normalized gain of the ageing part (A) to the gain of the reference part (B) along with the T/p as a function of accumulated charge per unit area is plotted in Figure~\ref{gain_charge}. No sign of ageing is observed after accumulation of more that 0.04~mC/mm$^{2}$. 

The energy resolution of the two parts as a function of time is shown in Figure~\ref{resotimeageing}. The energy resolution from both part varies between 16 to 19\% in the period of operation.

\section{Summary and Conclusion}\label{sec:summary}

In conclusion, systematic studies on the performances of the triple GEM detectors have been performed with a gas mixture of argon and CO$_{2}$ in 70/30 mixing ratio. The minimum ionizing particle (MIP) spectra has been taken for different GEM voltage settings and the spectra has been fitted with a Landau distribution. An efficiency plateau at 95\% has been achieved for cosmic ray. Crosstalk of the GEM signals are measured. About 10\% crosstalk has been observed between the two segments of the GEM.

At high rate operation of GEMs the value of the protection resistor influences the gain and the stability. This feature has been investigated varying both the rate and the value of the protection resistor, using both X-ray generator and Fe$^{55}$ sources. In both the measurements it is clear that the gain is decreasing with the rate. The rate of decrease is maximum with the maximum value of the protection resistor value.

The long-term stability and the ageing of triple GEM detectors has been studied employing both X-ray and Fe$^{55}$ sources. In the continuous operation of about 14 days the gain is measured and normalized with the T/p value. The normalized gain fluctuates with time for about a 10\% peak to peak value. This variation might comes from the variation of the  O$_2$ concentration in the gas as well as variation of the gas ratio due to changes of the characteristics of the mass flow controller with temperature. Further investigation is necessary to correlate this variation with other ambient parameters. The energy resolution varies about 4\% in this period of operation.

The ageing study of one GEM module is performed by using an 8 keV Cu X-ray generator to verify the stability and integrity of the GEM detectors over a longer period of time. The accumulated charge on the detector is calculated from the rate of the X-ray and the average gain of the detector. No sign of ageing is observed after accumulation of more that 0.04~mC/mm$^{2}$. Compared to the total accumulated charge of 7.5~C/cm$^{2}$ in the 10 years period of CBM experiment 0.004~C/cm$^{2}$ is just a first view into the ageing.

\acknowledgments

We are thankful to Dr. Ingo Fr\"{o}hlich of University of Frankfurt, Prof. Dr. Peter Fischer of Institut f\"{u}r Technische Informatik der Universit\"{a}t Heidelberg, Prof. Dr. Peter Senger, CBM Spokesperson and Dr. Subhasis Chattopadhyay, Deputy spokesperson, CBM  for their support in course of this work. We are also grateful to Dr. Leszek Ropelewski and Dr. Serge Duarte Pinto of RD51 for their valuable suggestions.

\end{document}